\documentclass[a4paper,
              11pt,
              accepted=2024-07-12,
              ]{quantumarticle}
\pdfoutput=1

\usepackage[numbers,sort&compress]{natbib}
\usepackage{amsmath}
\usepackage{amssymb}
\usepackage{amsthm}
    \newtheorem{theorem}{Theorem}
    \newtheorem{observation}[theorem]{Observation}
\usepackage{bbold}
\usepackage{bm}
\usepackage[colorlinks=true,
            linkcolor=quantumviolet,
            citecolor=quantumviolet,
            urlcolor=quantumviolet]{hyperref}
\usepackage{float}
\usepackage{graphicx}
	\graphicspath{{figs/}}
\usepackage{overpic}
\usepackage{mathtools}
\usepackage{xcolor}
  \definecolor{quantumviolet}{HTML}{53257F}
\usepackage[caption=false]{subfig}
\newcommand{\MPS}{\text{MPS}}

\makeatletter
\g@addto@macro{\UrlBreaks}{\do\/\do\-\do\_}
\makeatother

\usepackage[normalem]{ulem}

\newcommand{\removed}[1]{\textcolor{red}{\ifmmode\text{\sout{\ensuremath{#1}}}\else\sout{#1}\fi}}

\begin{document}

\title{Privacy-preserving machine learning with tensor networks}

\author{Alejandro Pozas-Kerstjens}
\affiliation{Group of Applied Physics, University of Geneva, 1211 Geneva 4, Switzerland}
\affiliation{Constructor Institute, 8200 Schaffhausen, Switzerland}
\affiliation{Instituto de Ciencias Matem\'aticas (CSIC-UAM-UC3M-UCM), 28049 Madrid, Spain}
\affiliation{Departamento de An\'alisis Matem\'atico, Universidad Complutense de Madrid, 28040 Madrid, Spain}
\author{Senaida Hern\'andez-Santana}
\affiliation{Departamento de Matem\'atica Aplicada a la Ingenier\'ia Industrial, Universidad Polit\'ecnica de Madrid, 28006 Madrid, Spain}
\author{Jos\'e Ram\'on Pareja Monturiol}
\affiliation{Instituto de Ciencias Matem\'aticas (CSIC-UAM-UC3M-UCM), 28049 Madrid, Spain}
\affiliation{Departamento de An\'alisis Matem\'atico, Universidad Complutense de Madrid, 28040 Madrid, Spain}
\author{Marco Castrill\'on L\'opez}
\affiliation{Departamento de \'Algebra, Geometr\'ia y Topolog\'ia, Universidad Complutense de Madrid, 28040 Madrid, Spain}
\author{Giannicola Scarpa}
\affiliation{Escuela T\'ecnica Superior de Ingenier\'ia de Sistemas Inform\'aticos, Universidad Polit\'ecnica de Madrid, 28031 Madrid, Spain}
\author{Carlos E. Gonz\'alez-Guill\'en}
\affiliation{Departamento de Matem\'atica Aplicada a la Ingenier\'ia Industrial, Universidad Polit\'ecnica de Madrid, 28006 Madrid, Spain}
\author{David P\'erez-Garc\'ia}
\affiliation{Instituto de Ciencias Matem\'aticas (CSIC-UAM-UC3M-UCM), 28049 Madrid, Spain}
\affiliation{Departamento de An\'alisis Matem\'atico, Universidad Complutense de Madrid, 28040 Madrid, Spain}

\begin{abstract}
    Tensor networks, widely used for providing efficient representations of low-energy states of local quantum many-body systems, have been recently proposed as machine learning architectures which could present advantages with respect to traditional ones.
    In this work we show that tensor network architectures have especially prospective properties for privacy-preserving machine learning, which is important in tasks such as the processing of medical records.
    First, we describe a new privacy vulnerability that is present in feedforward neural networks, illustrating it in synthetic and real-world datasets.
    Then, we develop well-defined conditions to guarantee robustness to such vulnerability,  which involve the characterization of models equivalent under gauge symmetry.
    We rigorously prove that such conditions are satisfied by tensor-network architectures.
    In doing so, we define a novel canonical form for matrix product states, which has a high degree of regularity and fixes the residual gauge that is left in the canonical forms based on singular value decompositions.
    We supplement the analytical findings with practical examples where matrix product states are trained on datasets of medical records, which show large reductions on the probability of an attacker extracting information about the training dataset from the model's parameters.
    Given the growing expertise in training tensor-network architectures, these results imply that one may not have to be forced to make a choice between accuracy in prediction and ensuring the privacy of the information processed.
\end{abstract}

\maketitle

\section{Introduction}
Vast amounts of data are routinely processed in machine learning pipelines, every time covering more aspects of our interactions with the world.
When the models processing the data are made public, is the safety of the data used for training it guaranteed?
This is a question of utmost importance when processing sensitive data such as medical records, but also for businesses whose competitive advantage lies in data quality.

The gold standard in privacy protection within machine learning~\cite{AppleDP,GoogleDP} is provided by differential privacy~\cite{DworkDP}, which consists of inserting carefully crafted noise either in the training dataset~\cite{warner1965,dwork2014}, in the final model parameters~\cite{DworkDP}, in the objective function~\cite{phan2017objective}, or in the gradient updates~\cite{tensorflowDP}, in order to hide the presence or absence of any particular sample in the training dataset.
There exist, however, privacy-related issues that do not directly fall in this category.
Imagine a machine-learning algorithm designed to diagnose a specific disease, which uses patients' records as training data.
Assume that these records have a strong imbalance in a particular morbidity, but it turns out that this morbidity has no association to the disease target of the model.
Even if irrelevant for the final task of the algorithm, knowing this imbalance may have consequences even at the individual level, if just the participation of a patient in the study (in contrast with the knowledge of their full record) is disclosed by other means.
As a first result, we show that this concern is a reality in machine learning architectures based on neural networks, caused by the driving of the corresponding network parameters to erase the information that is irrelevant.

In regards to the protection of privacy, the ideal model would only retain from the training dataset the information that is essential to perform well.
In this sense, access to the model's parameters in one such models would not be more informative about the data used for training than having a description of them by means of recording the outputs generated for different inputs.
This is a problem similar in spirit to the protection of software against Man-At-The-End attacks~\cite{MATE}.
As a second contribution, we show that the characterization of complex physical systems can provide a fruitful alternative viewpoint on the problem of privacy in machine learning.
Concretely, we rigorously prove that in specific tensor network architectures~\cite{verstraete2008tn, cirac2021matrix} ---a large family of architectures inspired by the entanglement structure of quantum many-body states--- it is possible and easy to find alternative parametrizations of a model that are as informative as a black-box access to it.
These architectures, known as matrix product states \cite{cirac2021matrix} (MPS), are promising learning architectures~\cite{stoudenmire,novikov2018} that now compete with \cite{stoudenmire2018,glasser2019probabilistic,selvan2020tensor} or even surpass traditional architectures in certain tasks, such as anomaly detection~\cite{VidalAnomaly}, sequence modelling~\cite{Miller2021sequence}, or generative modelling subject to constraints~\cite{LopezPiqueres2022}.
Moreover, the erasure of information irrelevant for the task at hand is achieved with no impact on the model's performance, in contrast with solutions based on differential privacy.
The core of our contribution is encompassed Observation \ref{theo:thm1} and Theorem \ref{theo:thm2}, which can be informally stated as
\begin{observation}[Informal version]
    If the sets of parameters that leave a model invariant can be characterized, then each white-box attack to a representative of the set is equally performing (in terms of accuracy) as an attack to the corresponding black box.
\end{observation}

\begin{theorem}[Informal version]
  There is a canonical form for the set of MPS architectures so that every white-box attack to such canonical-form set of parameters is ``as good'' (in terms of the attack accuracy and its regularity as a function, which characterizes how hard it is to perform the attack) as an attack to the black-box representation.
\end{theorem}

Notably, as part of the proof of Theorem~\ref{theo:thm2}, we construct a new canonical form for MPS that has a high degree of regularity and does not leave any residual gauge freedoms, in contrast with those based on singular value decompositions \cite{geometry}.

This work wants to drag focus to physics as a source of mathematically founded inspiration for machine learning solutions.
Cross-fertilizations between machine learning and physics are now commonplace~\cite{MLPhysReview}: on the one hand, machine learning algorithms are routinely used in particle colliders~\cite{radovic2018}, in the understanding of quantum matter and its properties~\cite{carrasquillaReview,rodriguez2019unsupervised}, or in the experimental control of quantum computers~\cite{niu2018control,fossel2018control}; on the other hand, tools developed within the umbrella of physics have proven invaluable in the understanding of the training and performance of machine learning algorithms, perhaps the most significant being the theory of the information bottleneck~\cite{tishby2000information}.
It is expected that such improved understanding leads to new proposals, inspired by physics, of machine learning architectures or training algorithms.
However, with the very notable exception of Boltzmann machines~\cite{nguyen2017BMreview,tramel2018tap,Pozas2021RAPID}, this type of influence of physics in machine learning has been arguably limited.
Our work is one of such form of influences, the main message being that tensor-network architectures provide a favorable framework to develop privacy-preserving machine learning algorithms. 

The article is organized as follows:
In Section \ref{sec:vulnerability} we mathematically describe the new privacy vulnerability present in neural networks and we illustrate it in toy and real-world datasets.
In Section \ref{sec:tns} we provide a broad introduction to the intuitions behind the requirements for robustness against the vulnerability and to the family of tensor networks known as matrix product states \cite{cirac2021matrix}, which will be used later on in the work.
Sections \ref{sec:thm1} and \ref{sec:thm2} are devoted to the formalization of the previous intuitions and to proving that the conditions are satisfied by tensor network architectures.
In Section \ref{sec:experiments} we show explicitly the privacy gains in a scenario where an attacker attempts to extract information about a training dataset of medical records from looking at the model's parameters.
Finally, we conclude in Section \ref{sec:discussion} with a discussion of concrete questions and research directions that this work opens.

\section{A new privacy vulnerability in feedforward neural networks}
\label{sec:vulnerability}
Training neural networks is routinely performed via optimization based on gradient descent: given a dataset to be learned and a parametrization of a family of models, a notion of error between the evaluation of the function on the dataset and the expected result is minimized by adjusting the parameters of the model in the direction given by the gradient of the error.
Also, most model classes have, at some point in their architecture, a concatenation of parametrized affine transformations, of the form $\bm{y}^{(l)}=W^{(l)}\cdot\bm{z}^{(l-1)}+\bm{b}^{(l)}$ where $W^{(l)}$ is a matrix of weights and $\bm{b}^{(l)}$ is a vector of biases, and fixed nonlinear functions, $\bm{z}^{(l)}=\phi_l(\bm{y}^{(l)})$, applied to the data.
These parameters, as we shall see now, contain information about the training dataset that, ideally, a model should not reveal.

\begin{figure}
  \centering
  \begin{minipage}[b]{\columnwidth}
    \subfloat[\label{fig:simple_before}]{
      \includegraphics[width=0.85\textwidth]{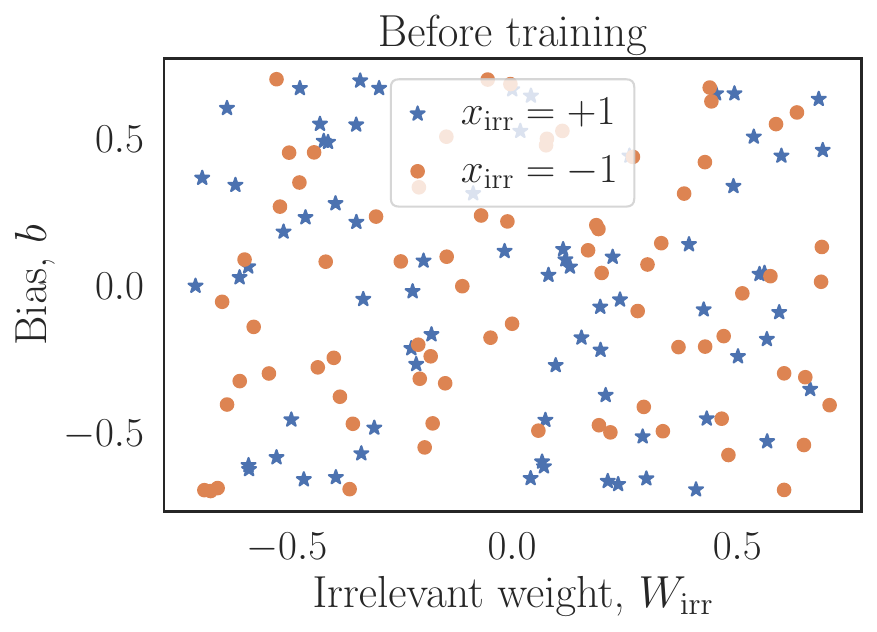}
	}
    \end{minipage}
  	\\
  	\begin{minipage}[b]{\columnwidth}
  	  \subfloat[\label{fig:simple_after}]{
    	\includegraphics[width=0.85\textwidth]{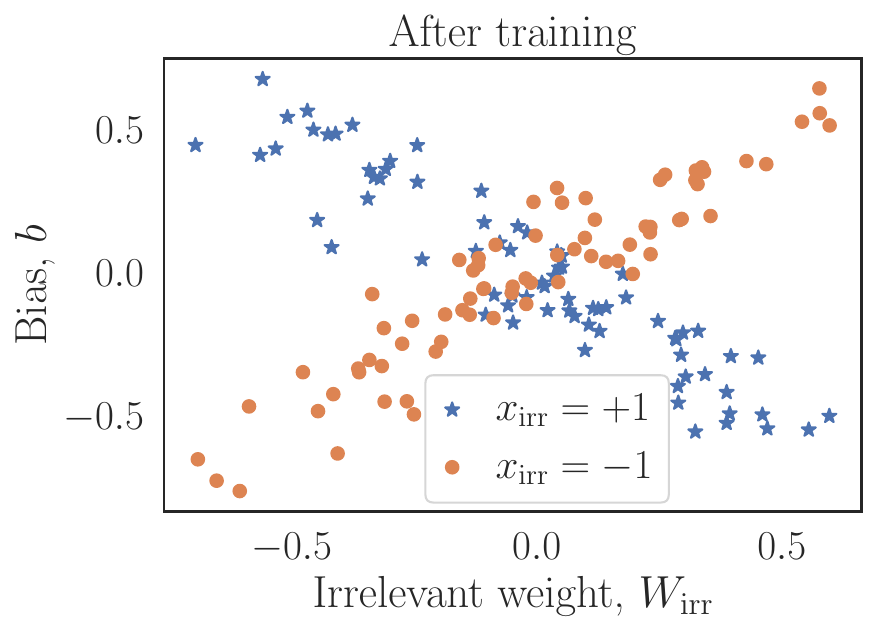}
      }
  \end{minipage}
  \caption{Illustration of the proposed vulnerability.
  Each point represents a simple neural network model, $f_\theta(\bm x)=\phi\left(W_\text{rel}x_\text{rel} + W_\text{irr}x_\text{irr} + b\right)$ (where $\phi$ is an activation function), that is trained to learn the function $y(\bm{x})=\text{sign}(x_\text{rel})$.
  Each model is trained on a different dataset where $x_\text{rel}\sim \mathcal{N}(0,1)$, and $x_\text{irr}$ is $+1$ for all datapoints used to train the models depicted by the blue stars and $-1$ for the orange circles.
  The plots show the values of the neural network weight for the irrelevant variable, $W_\text{irr}$, and the bias of the output neuron, $b$, \protect\subref{fig:simple_before} before and \protect\subref{fig:simple_after} after training on a different, random dataset for each model.
  In this architecture, the gradients of any loss function $\mathcal{L}$ are $\partial_{W_\text{irr}}\mathcal{L}=x_\text{irr}\phi'\partial_\phi\mathcal{L}$ and $\partial_{b}\mathcal{L}=\phi'\partial_\phi\mathcal{L}$, implying that $\partial_{W_\text{irr}}\mathcal{L}=x_\text{irr}\partial_{b}\mathcal{L}$.
  These gradients will naturally drive the parameters to distinguishable regimes, which can be identified after training as demonstrated in \protect\subref{fig:simple_after}.
  The codes for generating these figures are available in the computational appendix~\cite{compApp}.}
  \label{fig:simple}
\end{figure}

For simplicity, let us start considering a hypothetical situation (see Figure~\ref{fig:simple}) where all points in the dataset that we want to learn have one binary feature which takes always the same value, say $1$, and the model is a concatenation of these affine transformation and nonlinearities.
Since all the datapoints in the dataset have the same value of the binary feature, this information is of no use for the classification, and thus an ideal final model shall not depend on it.
During training, the contribution of the biases in the affine transformations are driven to compensate that of the corresponding weights, in the attempt of eliminating the effect of this irrelevant variable on the final prediction.
If the initial data has the constant value $1$, this will make that the corresponding weights and biases are directed towards taking opposite signs.
The situation is the opposite if the binary feature takes the constant value of $-1$.
In order to minimize the contribution of the irrelevant feature, in this case the gradient will drive the parameters towards taking values with same signs.
An attacker that has access to the model can thus easily recover the nature of the irrelevant feature just by looking at the parity of a set of weights and biases, as depicted in Fig.~\ref{fig:simple_after}.

In more realistic scenarios the situation is not as clear-cut\footnote{After all, if a feature takes the exact same value in the whole dataset, it is not reasonable to feed it to the model.}.
It is more common to have features which have some imbalance between the different values it can take.
In such situation, this type of vulnerability is still present, as we demonstrate in Figure~\ref{fig:attacks}.
There, we show an illustration with deep neural networks trained on real-world data of medical records derived from the \texttt{global.health} database~\cite{dataset} of COVID-19 cases around the world.
The task in which the neural networks are trained is the prediction of the outcome of the infection given demographics, symptoms, and the parity of the date when the case was recorded (this is the feature irrelevant to the task).
Then, the attacks follow the spirit of shadow training~\cite{ateniese2015hacking,shokri2017membership}: the attacker is provided with trained models and labels denoting the majority value of the irrelevant feature in the corresponding training set, and with them trains a meta-model that constitutes the attack.
Further details on the dataset and prediction task can be found in Appendix \ref{app:training}, and details on the attacks can be found in Appendix \ref{app:attacks}.
Notably, for the case of neural networks, simple logistic regression meta-models perform well in the attack task. Despite being performed by an attacker provided with non-realistic power, these results imply an important fact. Namely, that information about the training dataset that is irrelevant to the task performed by the neural network is learnt in the training process and stored in the network's parameters.

\section{Tensor networks as machine learning architectures with privacy guarantees}
\label{sec:tns}
The above example is a (rather extreme) illustration that neural-network architectures store information about the training set in the way that the network parameters process its features.
In the remainder of this work, we will see that in tensor networks~\cite{verstraete2008tn} this information can easily be deleted without compromising on model performance.

Continuing with the previous example, it is clear that there exists a straightforward way to erase the information about irrelevant features that does not lead to privacy leaks: simply setting to zero the weights that propagate the influence of those features to the initial layer of the network.
In a hypothetical situation where one fixed such parameters and trained the rest, the result would be an alternative collection of weights and biases leading to a model, ideally equally performing, yet not containing any information about the training dataset other than that needed for making the prediction.
Thus, the ability to characterize the sets of parameters that lead to the same model opens the door to ways of choosing model parameters which contain no more information about the training dataset than what can be inferred from recording the output for different inputs.
The first part of our contribution in this aspect is formally proving this intuition.
The second part is showing that for the family of matrix-product state architectures \cite{cirac2021matrix} (which have recently attracted interest in machine learning~\cite{stoudenmire,novikov2018}) one can make an assignment of parameters that has specially suitable properties when analyzed under the lens of privacy in machine learning.

\subsection{Matrix product states}
The field of quantum many-body physics has been developing manageable ways of simulating the states and evolutions of quantum systems composed of many particles.
The so-called tensor networks, introduced with the advent of quantum information theory as an architecture capable of capturing the entanglement content present in the low-energy sector of quantum many-body systems \cite{cirac2021matrix}, have recently gained attention in machine learning as alternative parametrizations of high-dimensional, convoluted functions \cite{stoudenmire, novikov2018, stoudenmire2018, glasser2019probabilistic, selvan2020tensor, VidalAnomaly}.
Matrix-product state (MPS) representations~\cite{davidMPS,VidalMPS} are the simplest and most widely used tensor network architecture when studying quantum many-body systems.
In fact, MPS were later rediscovered in the field of numerical analysis, under the name of tensor trains~\cite{oseledets09,oseledets}, which have been independently used for several applications in machine learning \cite{wahls2014,chen2018,kargas2021,wesel2021}.

In the context of machine learning, an MPS architecture is best defined when viewing complex functions as hyperplanes on high-dimensional feature maps of the input.
This is, when we consider (vector) functions $f(\bm{x})$ of the input $\bm{x}$ as taking the form
\begin{equation}
    f(\bm{x})=M\cdot\Psi(\bm{x}).
\end{equation}
MPS architectures correspond to the family of functions illustrated in Figure~\ref{fig:MPS}.
These are obtained when the vector feature map, $\Psi(\bm{x})$ (typically non-trainable), has a tensor-product form with one component per dimension of the input, $\Psi(\bm{x})=\psi_1(x_1)\otimes\psi_2(x_2)\otimes\dots\otimes\psi_N(x_N)$, and the hyperplane is expressed as a product of in general complex matrices (hence the name), namely
\begin{equation}
    \begin{aligned}
        M&{}^\ell _{s_1,s_2,\dots,s_N} \\
         &=\!\sum_{\{\alpha\}} [A_1]^{\alpha_1}_{s_1} [A_2]^{\alpha_1,\alpha_2}_{s_2} \cdots [A_j]^{\alpha_{j-1},\alpha_j}_\ell \cdots [A_N]^{\alpha_{N}}_{s_N},
    \end{aligned}
    \label{eq:mps}
\end{equation}
where $\ell$ is the free index in Fig.~\ref{fig:MPS} that stresses the vector character of $f(\bm{x})$.
These architectures have a very well-characterized gauge symmetry group, which determines the sets of parameters that describe the same function.
If between every two consecutive matrices one inserts a decomposition of the identity, $\mathbb{1}=Y_j\cdot Y_j^{-1}$ for an invertible matrix $Y_j$, the set of matrices given by $[B_j]_{s_j}=Y_j^{-1}\cdot [A_j]_{s_j}\cdot Y_{j+1}$ produce $M$ as well.
Importantly, it is well-known from quantum many-body physics that these are the only symmetries of the MPS architectures~\cite{davidMPS}, and that it is possible to fix a value of the gauge for each MPS.
This fixing is known as choosing a ``canonical form'', and it is generally obtained via singular value decompositions (SVD)~\cite{davidMPS,VidalMPS}.

\section{Privacy from reparametrization invariance}
\label{sec:thm1}
Invariance under reparametrizations is commonly known as gauge symmetry~\cite{gaugeBook}.
This is a concept that is commonplace in many-body physics, nuclear and particle physics, or in general relativity.
In order to rigorously prove that a complete characterization of gauge symmetries protects against revealing unintended information, we must first define some mathematical objects.
For a fixed learning architecture (this is, a functional ansatz depending on $n$ parameters), call $\mathcal{W}\in\mathbb{C}^n$ the set of all possible values of its parameters.
The architecture, along with a point $\theta\in\mathcal{W}$, completely determines a model.
Thus, we will refer to $\mathcal{W}$ as the set of white-box representations for a given architecture.
In this picture, training a model amounts to choosing the optimal $\theta$ for a specific task.
In general, this optimal value will depend on the data used for training the algorithm.
Analogously, the set of black-box representations $\mathcal{B}$ can be understood as the set of oracular functions (i.e., seen as input-output pairs) corresponding to each $\theta\in\mathcal{W}$.
In general, there exists a function $\pi:\mathcal{W}\rightarrow\mathcal{B}$ that assigns every white-box representation to its corresponding black-box oracle.
In certain cases one can define a right inverse, $\alpha:\mathcal{B}\rightarrow\mathcal{W}$, that assigns a set of parameters to a black-box representation.
This function satisfies $\pi\circ\alpha=\text{id}_{\mathcal{B}\rightarrow\mathcal{B}}$.
This is, the oracular function associated to a white-box representation of a black box is the black box itself.
Due to redundancies in the description, there can be many $\theta\in\mathcal{W}$ that lead to the same black-box representation, so in general, even if an $\alpha$ can be defined, it is not true that $\alpha\circ\pi=\text{id}_{\mathcal{W}\rightarrow\mathcal{W}}$.
A deterministic function $\alpha\circ\pi$ which takes all white boxes describing the same black box to the same element of $\mathcal{W}$ is commonly known as canonical form.

Consider also what an attack is.
In the proposed context, an attack is a function applied on a white-box representation whose output provides information about features of the training data.
Formally, we can model this as a (smooth) function $f:\mathcal{W}\rightarrow\mathbb{C}$ that maps the parameters of a model to a complex number.

With these, we can now state formally the idea behind the motivation of this work:
\renewcommand{\thetheorem}{1}
\begin{observation}
  Consider the set of white-box representations for some architecture, $\mathcal{W}\in\mathbb{C}^n$, and its associated set of black-box representations, $\mathcal{B}=\pi(\mathcal{W})$.
  If a right inverse $\alpha:\pi\circ\alpha=\text{id}_{\mathcal{B}\rightarrow\mathcal{B}}$ can be defined, then for every function $f$
  $$f(\hat{\theta})=\hat{f}[\pi(\hat{\theta})],$$
  where $\hat{\theta}=\alpha\circ\pi(\theta)\in\mathcal{W}$ are the parameters denoting a canonical form for the model described by $\theta$ and $\hat{f}=f\circ\alpha$ is the application of $f$ in black boxes.
  \label{theo:thm1}
\end{observation}
\begin{proof}
  By expanding $f(\hat{\theta})$:
  \begin{align*}
    f(\hat{\theta})&=f\circ\alpha\circ\pi(\theta)\\
    &=f\circ\alpha\circ\text{id}_{\mathcal{B}\rightarrow\mathcal{B}}\circ\pi(\theta)\\
    &=f\circ\alpha\circ\pi\circ\alpha\circ\pi(\theta)\\
    &=\hat{f}[\pi(\hat{\theta})],
  \end{align*}
  where we have decomposed the identity in the space of black boxes as \mbox{$\text{id}_{\mathcal{B}\rightarrow\mathcal{B}}=\pi\circ\alpha$}.
\end{proof}

The relation above means that the evaluation of an attack, $f$, in a set of parameters that describe the canonical form of a model, $\hat{\theta}=\alpha\circ\pi(\theta)$, coincides with the evaluation of the induced attack, $\hat{f}=f\circ\alpha$, in the associated black box, $\pi(\hat{\theta})=\pi(\theta)$.
Therefore, an attack $f$ cannot extract from a canonical form any information that is not present in the associated black-box.

Note that the only requirement of Observation~\ref{theo:thm1} is that the function $\alpha$ can be defined.
In particular, there are no requirements on the degree of regularity, or smoothness, of $\alpha$.
Thus, in general, the degree of regularity of the black-box attack $\hat{f}$ will depend on the regularity of both $f$ and $\alpha$.
In the next section we prove that, for the case of MPS, $\alpha$ can indeed be taken as smooth as possible: namely holomorphic.
This implies that the degree of regularity of the white-box and black-box attacks is the same, and therefore, that MPS are strong candidates for privacy-preserving machine learning architectures.

\section{Univocal canonical form in MPS}
\label{sec:thm2}
From the result in the previous section we know that if a canonical form can be defined, then the information about the training dataset stored in the model's parameters is the same information that is stored in the corresponding black box.
However, it could still be possible that extracting information present in the model is easier when having access to its parameters.
In this section we prove that this is not the case for MPS architectures.

It is important to note that the notion of canonical form defined in the previous section requires that all white boxes describing the same black box element are assigned univocally to \textit{the same} element of $\mathcal{W}$.
This already implies the existence of a function $\alpha:\mathcal{B}\rightarrow\mathcal{W}$ so that $\alpha\circ \pi:\mathcal{W}\rightarrow \mathcal{W}$ is the canonical form.
The known canonical forms for MPS do not assign the same white box element to all models equivalent under gauge symmetry.
For instance, the standard SVD-based canonical form of MPS \cite{davidMPS,VidalMPS} presents a residual $\mathrm{U}(1)\times\dots\times\mathrm{U}(1)$ gauge freedom~\cite{geometry}.
However, there are other decompositions of MPS that do not leave this residual freedom.

Thus, in the following we will show that a canonical form based on the skeleton decompositions of Ref.~\cite{oseledets2010decompo} is (i) univocal, so all gauge-equivalent models are mapped to the same canonical form, but also (ii) holomorphic, and hence possessing the highest degree of smoothness, and (iii) global, so the exact same procedure can be applied to any point in $\mathcal{W}$ (with the potential exception of subsets of measure zero). 
This will imply the same properties for the induced map $\alpha:\mathcal{B}\rightarrow \mathcal{W}$.
Then, since holomorphy implies that the composition of any function with $\alpha$ preserves its regularity, any attack at the white-box level, $f:\mathcal{W}\rightarrow \mathbb{C}$, can be upgraded to an attack at the black-box level, $\hat{f}=f\circ\alpha:\mathcal{B}\rightarrow \mathbb{C}$, with the same regularity.
Therefore, for every white-box attack to our canonical-form representation not only there exists a black-box attack with the same performance (as we showed in Observation~\ref{theo:thm1}), but also with the same regularity.
This is, for MPS, not only the canonical form stores as much information as the black box, but also extracting such information is equally hard in both cases.

Recall that MPS architectures process the input $\bm{x}$ via $f(\bm{x})=M\cdot\Psi(\bm{x})$, where $\Psi(\bm{x})$ is a (typically non-trainable) feature map with a tensor-product form and $M$ was described in Eq.~\eqref{eq:mps} with each entry of the $A$ tensors in the left-hand side being a trainable parameter.
Thus, when considering MPS architectures, the set $\mathcal{W}$ is a subset of $\mathbb{C}^n$ (where typically $n=Nb^2d$ for some number of sites $N$, bond dimension $b$, and physical dimension $d$)\footnote{See Refs.~\cite{cirac2021matrix,geometry} for descriptions of these quantities and their connections with the simulation of quantum many-body systems.} given by all the possible values of all the elements of the $A$ tensors in the right-hand side of Eq.~\eqref{eq:mps}.
The function $\pi$ that maps a white-box representation of an MPS to its corresponding black box is, precisely, Eq.~\eqref{eq:mps}.
Then, the set $\mathcal{B}$ is defined as $\pi(\mathcal{W})$, i.e., the subset of $\mathbb{C}^{n'}$ (with $n'=d^N$) defined by all allowed values of the entries of the $M$ tensor in the left-hand side of Eq.~\eqref{eq:mps}.

Since removing a zero-measure set will have no effect in the applicability of the properties of the canonical form (the matrices obtained after the training process will never belong to that set), we will consider a canonical form \textit{global} if $\mathcal{W}$ can be taken as the whole $\mathbb{C}^n$ except maybe for a proper, closed, smooth manifold of dimension strictly less than $n$, meaning in particular that $\mathcal{W}$ is open and its complementary set in $\mathbb{C}^n$ has measure zero.


The description of an univocal, global and holomorphic canonical form is done in the following theorem.
For simplicity we assume that the dimension of all legs in the tensors are the same.
However, it is easy to obtain a version of it with different dimensions.

\begin{widetext}
\renewcommand{\thetheorem}{2}
\begin{theorem}
    For an MPS defined by a collection of tensors $\{A_j\}_{j=1}^N$, a univocal, global, holomorphic canonical form is given by the skeleton decomposition defined by the tensors $\hat{A}_1=\mathbb{1}$, $\hat{A}_N=\mathbb{1}$, and $[\hat{A}_j]_{s_j}^{\beta_{j-1},\beta_j} = \sum_\gamma [L_j]_{s_j}^{\beta_{j-1},\gamma}[C_j]_{\gamma,\beta_j}$, where
  \begin{align*}
      [L_j]_{s_j}^{\beta_{j-1},\gamma}&=\sum_{\{\alpha\}} [A_1]^{\alpha_1}_{1} \cdots [A_{j-2}]^{\alpha_{j-3},\alpha_{j-2}}_{1}
      [B_j]^{\alpha_{j-2},\alpha_{j+1}}_{\beta_{j-1},s_j,\gamma}
      [A_{j+2}]^{\alpha_{j+1},\alpha_{j+2}}_{1} \cdots [A_N]^{\alpha_{N}}_{1},\\
      [B_j]^{\alpha_{j-2},\alpha_{j+1}}_{\beta_{j-1},s_j,\gamma} &= \sum_{\alpha_{j-1},\alpha_j}[A_{j-1}]^{\alpha_{j-2},\alpha_{j-1}}_{\beta_{j-1}}[A_j]^{\alpha_{j-1},\alpha_j}_{s_j}[A_{j+1}]^{\alpha_j,\alpha_{j+1}}_{\gamma},\\
      C_j&=(L_j)^{-1}\text{, with } [L_j]_{\gamma,\beta_j}=[L_j]_{\gamma}^{1,\beta_j}.
  \end{align*}
  This function induces a global, holomorphic map $\alpha:\mathcal{B}\rightarrow\mathcal{W}.$
  \label{theo:thm2}
\end{theorem}
\end{widetext}
Following the usual graphical notation for tensor networks (for its definition see, for instance, the explanations in Figure \ref{fig:results} or Ref.~\cite{cirac2021matrix}), the decomposition described in Theorem~\ref{theo:thm2} is
\begin{equation}
    \begin{minipage}[c]{0.18\columnwidth}
        \includegraphics{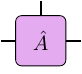}
    \end{minipage}
    \begin{minipage}[c]{0.04\columnwidth}
        \vspace*{0.1cm}=
    \end{minipage}
    \begin{minipage}[c]{0.33\columnwidth}
        \includegraphics{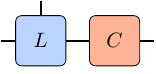}
    \end{minipage}
    \begin{minipage}[c]{0.05\columnwidth}
        \vspace*{0.15cm},
    \end{minipage}
    \label{eq:decomposition}
\end{equation}
with
\begin{align*}
    \begin{minipage}[c]{0.3\columnwidth}
        \includegraphics{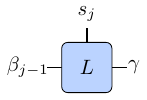}
    \end{minipage}
    &
    \begin{minipage}[t]{0.05\columnwidth}
        \vspace*{0.1cm}=
    \end{minipage}
    \begin{minipage}[c]{0.64\columnwidth}
        \includegraphics[scale=0.54]{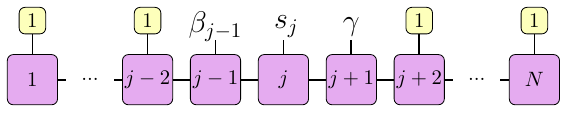}
    \end{minipage}
    \begin{minipage}[t]{0.05\columnwidth}
        \vspace*{0.2cm},
    \end{minipage}
    \\
    \begin{minipage}[c]{0.205\columnwidth}
        \includegraphics{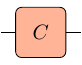}
    \end{minipage}
    &
    \begin{minipage}[c]{0.05\columnwidth}
        \vspace*{0.1cm}=
    \end{minipage}
    \Bigg(
    \begin{minipage}[c]{0.235\columnwidth}
        \includegraphics{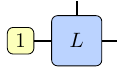}
    \end{minipage}
    \Bigg)^{-1},
    \\
    \begin{minipage}[c]{0.065\textwidth}
        \includegraphics{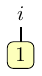}
    \end{minipage}
    &
    \begin{minipage}[c]{0.08\textwidth}
        \vspace*{0.1cm}=$\quad\,\,\delta_{i,1}$
    \end{minipage}.
\end{align*}

Now, let us proceed with the proof:
\begin{proof}
  The function $\{A\}\mapsto \{\hat{A}\}$ described is well defined whenever the $L$ matrices are invertible.
  When this is the case, the only non-trivial operation is the computation of such inverses.
  This is an holomorphic operation in the elements of the matrix.
  The matrices $L$, at every step in the procedure, are just a projection on specific sites of the original MPS.
  While the projected states are not invertible in all MPS, the set for which this is not the case is an algebraic variety of dimension strictly less than $n$ ($n$ being the total number of parameters).
  Note that this argument is independent of the particular projection chosen. This is, one could have chosen a projection different from $\delta_{i,1}$, and even different projections for different sites of the MPS, and the results would still hold.
  This shows that the canonical form is global.
  Moreover, standard algebraic calculations lead to verifying that $\pi(\{\hat{A}\})=\pi(\{A\})$. Note also that the canonical form obtained in this way will be the same for all MPS that are related by a gauge transformation, and hence univocal. This means that this canonical form indeed induces a well-defined map $\alpha: \mathcal{B}\rightarrow\mathcal{W}$ so that $\pi\circ\alpha$ is precisely the canonical form map $\{A\}\mapsto \{\hat{A}\}$.
  
  It thus remains to show that such $\alpha$ is also holomorphic.
  In order for this statement to be meaningful, or even to define what it means for an attack on black boxes to be smooth, one needs to endow the set $\mathcal{B}$ with a smooth structure; more specifically, with a complex manifold structure.
  Note that $\mathcal{W}$ is trivially a complex manifold, since it is an open subset of some $\mathbb{C}^n$.

  There are a priori two ways to see $\mathcal{B}$ as a complex manifold.
  One is as a submanifold, generated by the $\pi$ given by Eq.~\eqref{eq:mps} as $\mathcal{B}=\pi(\mathcal{W})$ and embedded in the ambient (exponentially large) complex vector space of all possible input-output relations.
  The other is as the space of orbits defined by the gauge symmetries.
  This is, if we call $\mathrm{S}_\MPS$ to the complex Lie group of gauge symmetries, the space of orbits is nothing but the quotient set $\mathcal{W}/\mathrm{S}_\MPS$, where a natural complex manifold structure can be defined whenever the symmetry group action is reasonably good (holomorphic, free and proper).
  One can see, using well-known results in the theory of complex manifolds, that both ways to describe $\mathcal{B}$ are equivalent.
  Formally, the complex manifolds $\mathcal{W}/\mathrm{S}_\MPS$ and $\mathcal{B}$ are biholomorphic, and thus we write $\mathcal{B}=\mathcal{W}/\mathrm{S}_\MPS$ from now onwards.
  The analysis of these manifolds has been done in full detail for the set of ``full-rank'' MPS in Ref.~\cite{geometry}, where we also refer to for the necessary definitions and background.
  Due to our additional condition of the matrices $L$ being invertible, our sets $\mathcal{W}$ and $\mathcal{B}$ are (full measure) subsets of those considered in Ref.~\cite{geometry}, and the exact same proof applies.

  By the way the complex manifold structure is defined in the space of orbits $\mathcal{W}/\mathrm{S}_\MPS$, for any holomorphic map  $\mathcal{W}\rightarrow \mathcal{W}$ that is invariant under the action of the gauge group $\mathrm{S}_\MPS$ ---as it is the case for the canonical form defined--- the uniquely defined associated map $\alpha:\mathcal{B}=\mathcal{W}/\mathrm{S}_\MPS\rightarrow {\mathcal{W}}$ is also holomorphic.
  All details can also be found in Ref.~\cite{geometry} and the references therein.
\end{proof}

Summarizing, we have proved that the skeleton decomposition of Ref.~\cite{oseledets2010decompo} leads to a univocal, holomorphic and global canonical form for MPS architectures.
The existence of such a canonical form implies that the MPS parameters obtained after computing the canonical form store no more information than the information already available in a black-box oracular access to the model, and that such information is equally hard to extract in both cases. 
Importantly, the sets of parameters generated by the gauge freedom all describe the exact same function, and thus the reparametrization to a canonical form does not have an impact on the performance of the model.
This is in stark contrast with approaches based on differential privacy to protect the privacy in neural networks, where noise is added to the training dataset or the final parameters, resulting in a tradeoff between the model’s utility and the protection of the dataset.

In particular, for the type of examples discussed earlier on features that are irrelevant to the target task, and in the ideal scenario in which the model learns perfectly, this means that the MPS parameters obtained after computing the canonical form would have no dependence at all on these irrelevant features.
This is indeed what is observed in realistic scenarios such as that illustrated in Figure~\ref{fig:attacks} when using the univocal canonical form described in Theorem~\ref{theo:thm2}, denoted there in the figure as MPS+U.

It is important to clarify that our Theorem~\ref{theo:thm2} is just a proof-of-principle illustration of the power of this approach to obtain privacy-preserving machine learning algorithms based on tensor networks.
We have focused on preserving regularity, but for any other property that one is interested in, in order to obtain a similar reduction from white boxes to black boxes, all that one needs is to find a canonical form that preserves such property.
In particular, it is important to note that the canonical form described in Theorem~\ref{theo:thm2} differs from that in the standard literature of MPS algorithms, based on a sequence of singular value decompositions \cite{VidalMPS,davidMPS}.
As stated earlier, the SVD-based canonical form of MPS presents a residual $\mathrm{U}(1)\times\dots\times\mathrm{U}(1)$ gauge freedom~\cite{geometry}, and it is not clear how to perform a global fixing of this residual gauge in an holomorphic manner.
Figure~\ref{fig:attacks} indicates that some information, albeit not as much as in the parameters obtained straight out from training, is still present in that freedom.
However, one can a priori expect at least some degree of privacy protection from the SVD-based canonical form.
Indeed, relaxing a bit the mechanism, the process of computing the canonical form of an MPS via SVD, and afterwards choosing randomly a representative of the residual gauge orbit, is equivalent to first computing the canonical form described in Theorem~\ref{theo:thm2}, and then computing the SVD-based canonical form of this MPS and randomly choosing a representative of the gauge orbit.
The latter process has all the privacy guarantees described in this work because of the application of the canonical form in Theorem~\ref{theo:thm2} in the first step, and therefore so does the former.
The red curve with pentagon-like markers denoted by MPS+C+S in Figure~\ref{fig:attacks} confirms that this argument also holds in numerical experiments, although the direct application of the univocal canonical form of Theorem~\ref{theo:thm2} leads to a lower variance.
This argument illustrates the power of our proof technique, since demonstrating that randomizing over the residual gauge after SVD is a procedure that makes it equally hard to extract information from white-box and black-box access requires to be able to analyze regularity properties of stochastic functions.

\section{Experiments}
\label{sec:experiments}
\begin{figure*}[ht!]
  \centering
  \begin{tabular}{cc}
    \begin{minipage}[t]{0.45\textwidth}
      \subfloat[\label{fig:NN}]{
        \includegraphics[width=0.9\textwidth]{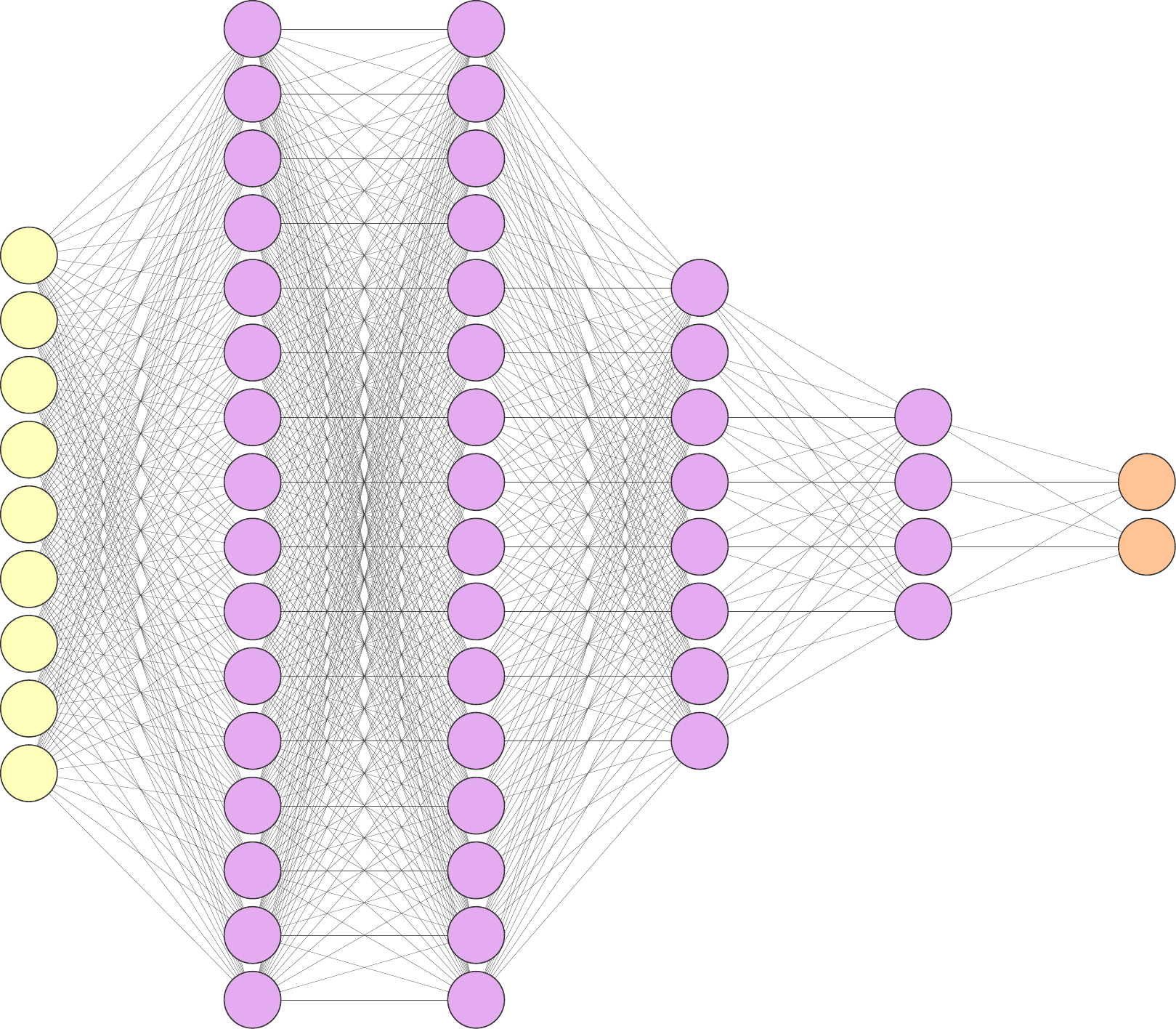}
      }
      \\
      \subfloat[\label{fig:MPS}]{
        \includegraphics[width=0.9\textwidth]{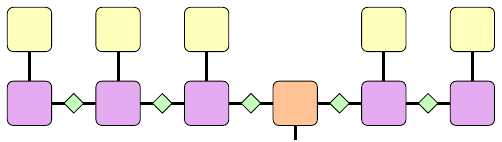}
      }
    \end{minipage}
    &
    \begin{minipage}[t]{0.45\textwidth}
      \subfloat[\label{fig:models}]{
        \includegraphics[width=0.85\textwidth]{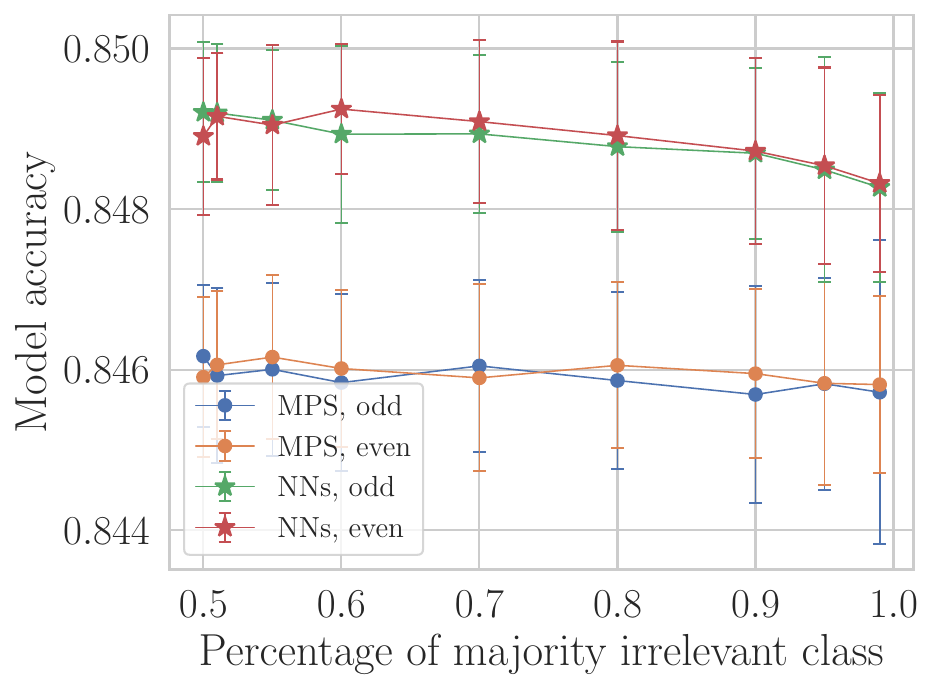}
       }
  	 \\
  	 \begin{minipage}[t]{\textwidth}
        \subfloat[\label{fig:attacks}]{
    	   \includegraphics[width=0.85\textwidth]{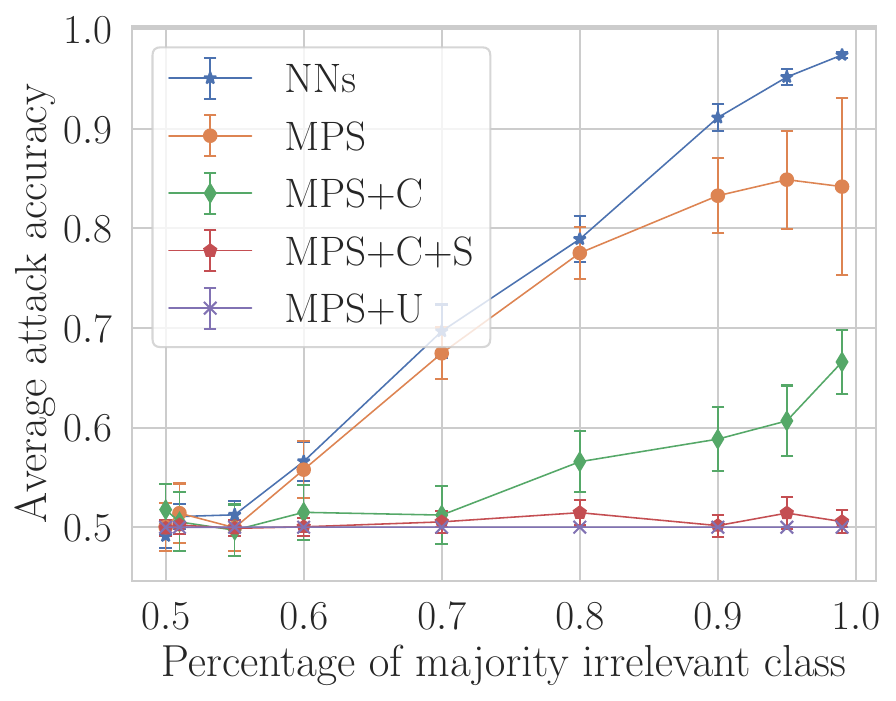}
        }
      \end{minipage}
    \end{minipage}
  \end{tabular}
  \caption{
    Comparison between deep neural networks and MPS architectures when learning a model predicting the outcome of COVID-19 infections given demographics and symptoms.
    Figures of merit are computed as a function of the percentage of dominant value in the irrelevant feature, namely the parity of the day of reporting.
    For every such percentage, several trainings are run for each of several different datasets, and statistics are computed over the full ensemble of resulting models.
    Figures \protect\subref{fig:NN} and \protect\subref{fig:MPS} show the neural network and MPS architecture used throughout the experiments.
    In \protect\subref{fig:MPS}, each element is a tensor with as many dimensions as legs.
    The purple squares in the lower row represent the parameters of the MPS.
    They are arranged in three-dimensional tensors, which are multiplied by their neighboring tensors and by the input.
    This is encoded in the one-dimensional vectors depicted by the yellow squares in the top row.
    The final tensor after multiplications via Eq.~\eqref{eq:mps} is a vector encoding the output, because the bottom leg of the orange square in the bottom row is free.
    The gauge symmetry that allows to erase information about irrelevant features is the decomposition of the identity, represented by the green diamonds, in invertible matrices that are later absorbed by the original tensors.
    Figures \protect\subref{fig:models} and \protect\subref{fig:attacks} show, respectively, the performance and vulnerability of neural networks and MPS trained on the COVID dataset, as a function of the imbalance between the two values of the irrelevant feature in the training set.
    The codes for generating these two figures are available in the computational appendix~\cite{compApp}.
    Figure~\protect\subref{fig:models} depicts the average accuracy in the whole database from which the different training sets are generated.
    The fact that models trained on datasets biased towards different values of the irrelevant feature perform equally indicates that the feature is indeed irrelevant.
    Figure~\protect\subref{fig:attacks} represents the accuracy of attacks attempting to predict the majority value of the irrelevant feature in the training dataset.
    Importantly, the MPS not expressed in canonical form are vulnerable in a similar way to the neural networks, and some information remains in the residual gauge freedom not fixed by the SVD-based canonical form (the green curve denoted by MPS+C). This information is erased when randomizing over the residual freedom (the red curve denoted by MPS+C+S), or by using the univocal canonical form of Theorem \ref{theo:thm2} (the purple curve denoted by MPS+U), which completely fixes the gauge.
    }
  \label{fig:results}
\end{figure*}
We illustrate the protective power of the canonical form of MPS in Figure~\ref{fig:results}.
Figure~\ref{fig:models} depicts the training of one of these MPS architectures in the real-world dataset used in the demonstration with neural networks (the prediction of the outcome of COVID-19 infections given demographics, symptoms, and the parity of the date of the record).
Figure~\ref{fig:attacks} shows the probability of success of attacks based on shadow training~\cite{ateniese2015hacking,shokri2017membership}, both before and after computing different canonical forms.
This sort of attacks provides upper bounds to the efficacy of realistic ones because it allows the attacker to have much more information than reasonable.
Details on the training and attacks can be found in appendices \ref{app:training} and \ref{app:attacks}, respectively.
Both types of models (neural networks and MPS architectures) perform very similarly in the target task (see Figure~\ref{fig:models}), while the vulnerability to attacks is markedly different: as can be seen in Figure~\ref{fig:attacks}, both neural networks and MPS architectures straight out from training are equally vulnerable, but MPS in canonical form are not.
For instance, at $80\%$ imbalance of the irrelevant feature, attacks on both models have around $78\%$ of accuracy, while in the canonical-form description of the MPS via SVD the accuracy drops down to around $56\%$, close to the limit of random guessing, but demonstrating that some information still remains in the residual gauge left by the SVD.
The limit of random guessing is achieved by randomly sampling the residual gauge freedom (this is the curve denoted by MPS+C+S) via adding decompositions of the identity in the form of diagonal matrices with random entries in $\{-1,1\}$, or by using the univocal canonical form described in Theorem~\ref{theo:thm2} (corresponding to the curve denoted by MPS+U).
Notably, as described in Appendix \ref{app:attacks}, the protection happens despite the attacks performed to MPS architectures being more general than those performed to neural networks: while linear regression models were able to extract the information on the irrelevant feature in the case of neural networks, in the case of MPS architectures the accuracy of the extraction is not matched even when using more powerful, deep neural networks as attack models.

One could be tempted to attribute the difference in robustness to the attacks to the difference in number of parameters of the models.
As explained in Appendix~\ref{app:training}, the neural network models contain 614 trainable parameters, while the MPS models contain only 40.
This could, in principle, lead to neural network models having a larger capacity that makes them more vulnerable.
However, the line corresponding to MPS models without post-processing (orange circles) in Fig.~\ref{fig:attacks} demonstrates that this is not the case: if the parameters of the MPS models are not brought to a canonical form, the models are vulnerable to the attacks in a way analogous to the neural network models.

\section{Discussion}
\label{sec:discussion}
As machine learning permeates through more layers in society, it is increasingly important to shift the focus from prediction accuracy to greater goals such as privacy and fairness.
This work shows that global information about the training dataset is hidden in the parameters of deep learning models, even if this information is irrelevant for the task at hand.
More importantly, it points at physics, and more concretely at the tensor networks used in quantum many-body physics, as a favourable framework where to find architectures that are robust to privacy leaks.
In this second aspect, our results are encompassed in Observation \ref{theo:thm1} and Theorem \ref{theo:thm2}.



We have shown that the key of the protection comes from the ability to characterize all sets of parameters that produce the same trained model.
Once this set is characterized, making the final choice of parameters in a way that is independent of the training dataset (i.e., choosing a canonical form) implies that the final parameters contain no more information about the training dataset than the strictly necessary to produce the output.
For neural networks, one could imagine obtaining a canonical form by proceeding in the spirit of model extraction attacks~\cite{tramer2016stealing,jagielski2020extraction}.
While one would need to study carefully whether the requirements for robustness are satisfied in this scenario, one should bear in mind the increased computation time and amount of necessary data (after all, a second model must be trained from input-output queries to the vulnerable one, ideally using datapoints not used in the training of the first), and the potential loss of accuracy entailed by the process.

The standard in privacy protection within machine learning is provided by differential privacy.
While now having become the standard in the industry in regards to privacy-preserving machine learning~\cite{AppleDP,GoogleDP}, adding the noise needed for ensuring differential privacy imposes a tradeoff between the level of privacy and the utility of the model.
Moreover, the vulnerability illustrated regards global properties of the dataset rather than individual datapoints.
Therefore, whether differential privacy protects against the vulnerability demonstrated remains to be understood.

With this work we want to shift the focus towards a promising class of architectures for machine learning.
Our contribution is a foundational step in that direction, where many open questions still lie ahead.
We have provided only an exemplary family of architectures, that of matrix product states, which we prove to admit parametrizations that are no more informative than a black-box access to a trained model.
Our results nevertheless apply to any architecture where different parametrizations lead to the same model and where a global, smooth, and one-to-one mapping can be defined between the space of black-box representations and a representative of every possible model.
If such mapping exists, one can use it to eliminate information that the final model should not have.
This part of the proof is not restricted neither to MPS architectures, to general tensor-network architectures, nor even to neural networks.
However, given all the expertise of the community of quantum many-body physics, it is expected that model classes that satisfy the required conditions are easier to find within the tensor-network family~\cite{molnarpeps}.
Finding such architecture classes, which are also enough expressive and easy to train, will constitute a very important task.
A promising family is that of canonical polyadic decompositions of tensors, which are commonplace architectures \cite{kargas2021,wesel2021} that are known to be invariant under suitable rescalings and for which, under mild conditions, a unique factorization exists \cite{sidiropoulos2000}.

In contrast with defenses based on differential privacy, the canonical form guarantees that defense in tensor-network architectures can be achieved with comparably little computational overhead (namely only that required for computing the canonical form) and without compromising on prediction accuracy.
However, tensor network architectures can also be trained using differentially private mechanisms, thus adding privacy of individual datapoints on top of the canonical form.
Still, more effort must be put into the optimization of training of tensor-network architectures, which only recently are finding advantages over state-of-the-art deep neural network architectures~\cite{VidalAnomaly,Miller2021sequence,LopezPiqueres2022}.
For this, it is fundamental to develop tensor-network models beyond the highly structured ones developed within the physics community (which, nevertheless, already cover large families of interesting architectures~\cite{molnarpeps}).
A good source of inspiration is the ``tensorization'' of popular deep learning models, a field that is rapidly gaining traction~\cite{Liu_2019,tensorLSTM,tensortransformer}.
More broadly, both neural-network and tensor-network architectures can easily be combined~\cite{glasser2019probabilistic,Kuznetsov2019TRIP,cheng2021PEPS}, so it is not unreasonable to imagine having hybrid architectures where tensor-network layers take care of some privacy aspects, and neural-network layers do the heavy-lifting on private data.

Going beyond machine learning, the equivalence between black-box and white-box access to functions described by MPS that we have proven in this work has also the potential of having important impacts in cryptography.
Indeed, equivalence between black and white boxes is the fundamental concept in obfuscation~\cite{barak2001obfuscation}.
Usually discussed in the context of programs or circuits, obfuscation refers to providing an alternative object that performs the same task but whose description is ``unintelligible'', meaning that anything that could be understood from this description could as well be understood just from input-output behavior.
The line of argument drawn by Observation \ref{theo:thm1} and Theorem \ref{theo:thm2} follows a spirit very similar to a proof of obfuscation, motivating a deeper exploration of the relations between tensor networks and Boolean circuits, and describing a path for proofs of obfuscation in restricted classes of circuits.

\begin{acknowledgments}
    We thank Thomas Vidick, Juan Jos\'e Garc\'ia-Ripoll, Sofyan Iblisdir, J. Ignacio Cirac and Esther Cruz for discussions and comments. 
    This work is supported by the European Union (Horizon 2020 research and innovation programme-grant agreement No. 648913 and ERDF ``A way of making Europe''), the Spanish Ministry of Science and Innovation (``Severo Ochoa Programme for Centres of Excellence in R\&D'' CEX2019-000904-S and ICMAT Severo Ochoa project SEV-2015-0554, and grants CEX2019-000904-S-20-4, MTM2014-54240-P, MTM2017-88385-P, PGC2018-098321-B-I00 and PID2020-113523GB-I00), the Spanish Ministry of Economic Affairs and Digital Transformation (project QUANTUM ENIA, as part of the Recovery, Transformation and Resilience Plan, funded by EU program NextGenerationEU), the Spanish Ministry of Universities (funded by EU program NextGenerationEU), Comunidad de Madrid (PEJ-2021-AI/TIC-23267 and QUITEMAD-CM P2018/TCS-4342), and the CSIC Quantum Technologies Platform PTI-001.
\end{acknowledgments}

\appendix

\section{Dataset and training of models}
\label{app:training}

\begin{table*}[t]
    \centering
    \begin{tabular}{c|c|c|c|c|c|c}
        & parity & country & age & gender & symptoms & recovery \\
        \hline\hline
        parity & $1$ & $0.005$ & $0.001$ & $0.001$ & $-0.012$ & $-0.002$ \\
        country & & $1$ & $0.059$ & $0.033$ & $0.161$ & $-0.055$ \\
        age & & & $1$ & $-0.010$ & $0.128$ & $-0.707$ \\
        gender & & & & $1$ & $0.004$ & $0.078$ \\
        symptoms & & & & & $1$ & $-0.143$ \\
        recovery & & & & & & $1$
    \end{tabular}
    \caption{Pearson correlation coefficients between the columns of the dataset used for training the neural network and MPS models.
    The columns in the table correspond to the columns (in order)  \texttt{events.confirmed.date}, \texttt{location.country}, \texttt{demographics.ageRange.start},  \texttt{demographics.gender}, \texttt{symptoms.status}, and \texttt{events.outcome.value} of the original database.}
    \label{tab:corrs}
\end{table*}

To illustrate the fact that neural networks are vulnerable to irrelevant feature leaks and MPS are not, we have trained both architectures in a real-world dataset, that is part of the \texttt{global.health} database of COVID-19 cases~\cite{dataset}.
This is a very large database that allows us to make small partitions, so that we can train shadow models when illustrating the attacks.
We take the dataset available on March 22nd 2021, use the data of two countries, Argentina and Colombia, for generating our database.
The database built only contains the columns \texttt{location.country}, \texttt{events.outcome.value}, \texttt{events.confirmed.date}, \texttt{demographics.ageRange.start}, \texttt{demographics.gender}, and \texttt{symptoms.status} from the whole dataset, and all datapoints where at least one of these entries is empty are discarded.
As a first balancing between countries, we take all entries for Argentina and, evenly, one out of every seven (which is approximately the ratio between the number of datapoints for each country) points for Colombia's cases where the column \texttt{events.outcome.value} takes the value \texttt{Death}, and the same amount for the cases where \texttt{events.outcome.value} takes any other value.
As a second balancing, now between cases with odd and even registration dates, we take all cases where the column \texttt{events.outcome.value} takes the value \texttt{Death} (a total of \mbox{21 503}), and the first \mbox{5 375} cases for each combination of country and parity from the subset of points where the column \texttt{events.outcome.value} takes the value \texttt{Recovered}. 
We provide the computer codes, written in Python, for generating the database from the global.health dataset in the computational appendix~\cite{compApp}.

The classification task in which both, neural networks and MPS, are trained, is predicting the value for \texttt{events.outcome.value} when provided the rest.
While all features are discrete in nature, we treat the age feature as continuous.
As irrelevant feature, we choose the parity of the report date, extracted from \texttt{events.confirmed.date}.
While this quantity is indeed expected to be irrelevant for the classification, we check that it is sufficiently uncorrelated with the remaining features (see Table~\ref{tab:corrs}).

The neural network model used is depicted in Figure~\ref{fig:NN} and consists of a five-layer architecture with structure 16-16-8-4-2, totalling 614 trainable parameters.
Each intermediate layer has a rectified linear unit as activation.
Training optimizes the cross-entropy between the predictions and the labels in batches of 8 datapoints, using the Adam optimizer with learning rate of $3\cdot10^{-4}$ and $\ell_2$ regularization of magnitude $6\cdot10^{-3}$, for at most \mbox{1 250} epochs.
The final model picked is that along the training history that achieves higher accuracy in a held-out validation set composed of \mbox{5 000} samples of the original database.

The matrix product state model is depicted in Figure~\ref{fig:MPS} and consists of six tensors where all the dimensions have cardinality $2$.
This is, the leftmost and rightmost purple squares in Figure~\ref{fig:MPS} are $2\times 2$ matrices, and the remaining purple squares and the orange square are tensors of dimensions $2\times 2\times 2$.
All entries in all tensors (a total of 40) are free, trainable parameters.
Training optimizes the cross-entropy between the predictions and the labels in batches of 100 datapoints, using the Adam optimizer with learning rate of $10^{-1}$, for 20 epochs.

The encoding of categorical variables in the input is different for both architectures.
For neural networks we perform a traditional one-hot encoding.
In contrast, for MPS we perform a noisy encoding, in such a way that for each datapoint the class is encoded in a random value in either $\left[0, \frac{1}{2}-\epsilon\right]$ or $\left[\frac{1}{2}+\epsilon, 1\right]$, with $\epsilon=5\cdot10^{-2}$.
Then, every dimension of the input is encoded in a different two-dimensional vector via $\psi(x) = (1-x, x)$, and these are the objects that are input to the MPS (the yellow squares in Figure~\ref{fig:MPS}).
For the \texttt{demographics.ageRange.start} column, a min-max normalization is performed before producing the input vector.
The different encoding of the variables for neural networks and MPS does not have severe impacts in the vulnerability of the models: as demonstrated in Figure~\ref{fig:attacks}, the MPS architectures are still vulnerable if no protection is applied.

The computation of representatives of the orbits of parametrizations of a same model is done in three different ways: via singular value decompositions (the green curve, denoted by MPS+C in Fig.~\ref{fig:attacks}), sampling uniformly over the residual gauge freedom after computing the singular value decomposition (the red curve, denoted by MPS+C+S), and using the prescription developed in Theorem \ref{theo:thm2} (the purple curve, denoted by MPS+U).
As we commented in Section \ref{sec:thm2}, it is left for future work to analyse whether the SVD-based canonical form fulfills similar regularity properties as the new one we construct here for the proof of Theorem \ref{theo:thm2}.
In any case, the analysis on Figure~\ref{fig:results} indicates that, in practice, the canonical form obtained via SVD guarantees a significant degree of privacy, although some information is still encoded in the residual gauge freedom left by the SVD that can be erased via randomization.

\section{Attacks}
\label{app:attacks}
The attacks we consider are in the spirit of shadow modeling attacks~\cite{ateniese2015hacking,shokri2017membership}, whereby the attacker is provided with a large collection of models trained on datasets that share the statistics of the dataset where the victim model has been trained on.
This is a situation where the attacker has much more power than what is realistic, so the results are upper bounds to realistic attacks.

In order to produce Figure~\ref{fig:attacks}, we generate, for every percentage of the majority class in the irrelevant variable, a total of 200 datasets (100 for each majority class) with that proportion, randomly sampled from the original dataset.
On each of these datasets, a total of 100 models are trained according to the prescription described in the previous section.
This produces, for each of the percentages (the horizontal axis of Figure~\ref{fig:attacks}), \mbox{20 000} trained models.
Out of all of them, those corresponding to 80 of the datasets are provided to the attacker along with their corresponding majority class, while the models for the remaining 20 datasets will be those to be attacked.

The attacker, depending on the type of models, does a different kind of attack: for neural networks the attack consists of a logistic regressor over the full set of models' weights and biases after a proper normalization, with $\ell_2$ regularization and using the LBFGS solver.
This attack is arguably simple, but this only reinforces the argument that neural networks are very vulnerable to leaking the nature of irrelevant features.
Removing parameters and keeping those in the initial layer, or doing PCA to reduce dimensionality, did not offer advantages to using the full set of model parameters as input.

For MPS architectures, the attacker trains a deep feedforward neural network that is input the model weights (after a proper normalization) and outputs the corresponding label.
This was done to provide the attack with more generality, so it could hopefully extract more information from the parameters of the model.
The attack neural network consists of six layers with structure 20-20-10-10-2-2, using rectified linear units as activation functions for the intermediate layers.
Training optimizes the cross-entropy between the predictions and the labels in batches of \mbox{1 000} datapoints, using the Adam optimizer with learning rate of $10^{-3}$ and $\ell_2$ regularization of magnitude $10^{-4}$, for at most \mbox{1 000} epochs.
The set of 80 datasets is split randomly in an 80-20 proportion as to generate a validation set for early stopping.

In order to provide statistics, we train and evaluate the attacks for several random choices of the 80 datasets given to the attacker.
We do a total of \mbox{1 000} of these repetitions, which provide the confidence intervals in Figure~\ref{fig:results}.

\bibliographystyle{quantum}
\bibliography{bibliography}

\begin{thebibliography}{10}

\bibitem{AppleDP}
Apple.
\newblock ``Differential privacy overview''.
\newblock
  \url{https://www.apple.com/privacy/docs/Differential_Privacy_Overview.pdf}~(2021).
\newblock Accessed: 2021-12-02.

\bibitem{GoogleDP}
Google.
\newblock ``How we’re helping developers with differential privacy''.
\newblock
  \url{https://developers.googleblog.com/2021/01/how-were-helping-developers-with-differential-privacy.html}~(2021).
\newblock Accessed: 2021-12-02.

\bibitem{DworkDP}
Cynthia Dwork, Frank McSherry, Kobbi Nissim, and Adam Smith.
\newblock ``Calibrating noise to sensitivity in private data analysis''.
\newblock \href{https://dx.doi.org/10.29012/jpc.v7i3.405}{J. Priv. Confid. {\bf
  7}, 17--–51}~(2017).

\bibitem{warner1965}
Stanley~L. Warner.
\newblock ``Randomized response: A survey technique for eliminating evasive
  answer bias''.
\newblock \href{https://dx.doi.org/10.2307/2283137}{J. Am. Stat. Assoc. {\bf
  60}, 63--69}~(1965).

\bibitem{dwork2014}
Cynthia Dwork and Aaron Roth.
\newblock ``The algorithmic foundations of differential privacy''.
\newblock \href{https://dx.doi.org/10.1561/0400000042}{Found. Trends Theor.
  Comput. Sci. {\bf 9}, 211--407}~(2014).

\bibitem{phan2017objective}
NatHai Phan, Xintao Wu, and Dejing Dou.
\newblock ``Preserving differential privacy in convolutional deep belief
  networks''.
\newblock \href{https://dx.doi.org/10.1007/s10994-017-5656-2}{Mach. Learn. {\bf
  106}, 1681--1704}~(2017).
\newblock  \href{http://arxiv.org/abs/1706.08839}{arXiv:1706.08839}.

\bibitem{tensorflowDP}
Martin Abadi, Andy Chu, Ian Goodfellow, H.~Brendan McMahan, Ilya Mironov, Kunal
  Talwar, and Li~Zhang.
\newblock ``Deep learning with differential privacy''.
\newblock In Proceedings of the 2016 ACM SIGSAC Conference on Computer and
  Communications Security.
\newblock \href{https://dx.doi.org/10.1145/2976749.2978318}{Pages 308--318}.
\newblock CCS '16New York, NY, USA~(2016). Association for Computing Machinery.
\newblock  \href{http://arxiv.org/abs/1607.00133}{arXiv:1607.00133}.

\bibitem{MATE}
Christian Collberg, Jack Davidson, Roberto Giacobazzi, Yuan~Xiang Gu, Amir
  Herzberg, and Fei-Yue Wang.
\newblock ``Toward digital asset protection''.
\newblock \href{https://dx.doi.org/10.1109/MIS.2011.106}{IEEE Intell. Syst.
  {\bf 26}, 8--13}~(2011).

\bibitem{verstraete2008tn}
Frank Verstraete, Valentin Murg, and J.~Ignacio Cirac.
\newblock ``Matrix product states, projected entangled pair states, and
  variational renormalization group methods for quantum spin systems''.
\newblock \href{https://dx.doi.org/10.1080/14789940801912366}{Adv. Phys. {\bf
  57}, 143--224}~(2008).
\newblock  \href{http://arxiv.org/abs/0907.2796}{arXiv:0907.2796}.

\bibitem{cirac2021matrix}
J.~Ignacio Cirac, David P\'erez-Garc\'{\i}a, Norbert Schuch, and Frank
  Verstraete.
\newblock ``Matrix product states and projected entangled pair states:
  Concepts, symmetries, theorems''.
\newblock \href{https://dx.doi.org/10.1103/RevModPhys.93.045003}{Rev. Mod.
  Phys. {\bf 93}, 045003}~(2021).
\newblock  \href{http://arxiv.org/abs/2011.12127}{arXiv:2011.12127}.

\bibitem{stoudenmire}
E.~Miles Stoudenmire and David~J. Schwab.
\newblock ``Supervised learning with tensor networks''.
\newblock In
  \href{https://proceedings.neurips.cc/paper/2016/hash/\\5314b9674c86e3f9d1ba25ef9bb32895-Abstract.html}{Advances
  in Neural Information Processing Systems}.
\newblock Volume~29, pages 4799--4807.
\newblock Curran Associates, Inc.~(2016).
\newblock  \href{http://arxiv.org/abs/1605.05775}{arXiv:1605.05775}.

\bibitem{novikov2018}
Alexander Novikov, Mikhail Trofimov, and Ivan~V. Oseledets.
\newblock ``Exponential machines''.
\newblock \href{https://dx.doi.org/10.24425/bpas.2018.125926}{Bull. Pol. Acad.
  Sci.: Tech. Sci. {\bf 66}, 789--797}~(2018).
\newblock  \href{http://arxiv.org/abs/1605.03795}{arXiv:1605.03795}.

\bibitem{stoudenmire2018}
E.~Miles Stoudenmire.
\newblock ``Learning relevant features of data with multi-scale tensor
  networks''.
\newblock \href{https://dx.doi.org/10.1088/2058-9565/aaba1a}{Quantum Sci.
  Technol. {\bf 3}, 034003}~(2018).
\newblock  \href{http://arxiv.org/abs/1801.00315}{arXiv:1801.00315}.

\bibitem{glasser2019probabilistic}
Ivan Glasser, Nicola Pancotti, and J.~Ignacio Cirac.
\newblock ``From probabilistic graphical models to generalized tensor networks
  for supervised learning''.
\newblock \href{https://dx.doi.org/10.1109/ACCESS.2020.2986279}{IEEE Access
  {\bf 8}, 68169--68182}~(2020).
\newblock  \href{http://arxiv.org/abs/1806.05964}{arXiv:1806.05964}.

\bibitem{selvan2020tensor}
Raghavendra Selvan and Erik~B. Dam.
\newblock ``Tensor networks for medical image classification''.
\newblock In
  \href{https://proceedings.mlr.press/v121/selvan20a.html}{Proceedings of the
  Third Conference on Medical Imaging with Deep Learning}.
\newblock Volume 121 of Proceedings of Machine Learning Research, pages
  721--732.
\newblock PMLR~(2020).
\newblock  \href{http://arxiv.org/abs/2004.10076}{arXiv:2004.10076}.

\bibitem{VidalAnomaly}
Jinhui Wang, Chase Roberts, Guifre Vidal, and Stefan Leichenauer.
\newblock ``Anomaly detection with tensor networks''~(2020).
\newblock  \href{http://arxiv.org/abs/2006.02516}{arXiv:2006.02516}.

\bibitem{Miller2021sequence}
Jacob Miller, Guillaume Rabusseau, and John Terilla.
\newblock ``Tensor networks for probabilistic sequence modeling''.
\newblock In
  \href{https://proceedings.mlr.press/v130/miller21a.html}{Proceedings of The
  24th International Conference on Artificial Intelligence and Statistics}.
\newblock Volume 130 of Proceedings of Machine Learning Research, pages
  3079--3087.
\newblock PMLR~(2021).
\newblock  \href{http://arxiv.org/abs/2003.01039}{arXiv:2003.01039}.

\bibitem{LopezPiqueres2022}
Javier Lopez-Piqueres, Jing Chen, and Alejandro Perdomo-Ortiz.
\newblock ``Symmetric tensor networks for generative modeling and constrained
  combinatorial optimization''.
\newblock \href{https://dx.doi.org/10.1088/2632-2153/ace0f5}{Mach. Learn.: Sci.
  Technol. {\bf 4}, 035009}~(2023).
\newblock  \href{http://arxiv.org/abs/2211.09121}{arXiv:2211.09121}.

\bibitem{geometry}
Jutho Haegeman, Michaël Mariën, Tobias~J. Osborne, and Frank Verstraete.
\newblock ``Geometry of matrix product states: Metric, parallel transport, and
  curvature''.
\newblock \href{https://dx.doi.org/10.1063/1.4862851}{J. Math. Phys. {\bf 55},
  021902}~(2014).
\newblock  \href{http://arxiv.org/abs/1210.7710}{arXiv:1210.7710}.

\bibitem{MLPhysReview}
Giuseppe Carleo, Ignacio Cirac, Kyle Cranmer, Laurent Daudet, Maria Schuld,
  Naftali Tishby, Leslie Vogt-Maranto, and Lenka Zdeborov\'a.
\newblock ``Machine learning and the physical sciences''.
\newblock \href{https://dx.doi.org/10.1103/RevModPhys.91.045002}{Rev. Mod.
  Phys. {\bf 91}, 045002}~(2019).
\newblock  \href{http://arxiv.org/abs/1903.10563}{arXiv:1903.10563}.

\bibitem{radovic2018}
Alexander Radovic, Mike Williams, David Rousseau, Michael Kagan, Daniele
  Bonacorsi, Alexander Himmel, Adam Aurisano, Kazuhiro Terao, and Taritree
  Wongjirad.
\newblock ``Machine learning at the energy and intensity frontiers of particle
  physics''.
\newblock \href{https://dx.doi.org/10.1038/s41586-018-0361-2}{Nature {\bf 560},
  41--48}~(2018).

\bibitem{carrasquillaReview}
Juan Carrasquilla.
\newblock ``Machine learning for quantum matter''.
\newblock \href{https://dx.doi.org/10.1080/23746149.2020.1797528}{Adv. Phys.: X
  {\bf 5}, 1797528}~(2020).
\newblock  \href{http://arxiv.org/abs/2003.11040}{arXiv:2003.11040}.

\bibitem{rodriguez2019unsupervised}
Joaquin~F. Rodriguez-Nieva and Mathias~S. Scheurer.
\newblock ``Identifying topological order through unsupervised machine
  learning''.
\newblock \href{https://dx.doi.org/10.1038/s41567-019-0512-x}{Nat. Phys. {\bf
  15}, 790--795}~(2019).

\bibitem{niu2018control}
Murphy~Yuezhen Niu, Sergio Boixo, Vadim Smelyanskiy, and Hartmut Neven.
\newblock ``Universal quantum control through deep reinforcement learning''.
\newblock \href{https://dx.doi.org/10.1038/s41534-019-0141-3}{npj Quantum Inf.
  {\bf 5}, 33}~(2019).
\newblock  \href{http://arxiv.org/abs/1803.01857}{arXiv:1803.01857}.

\bibitem{fossel2018control}
Thomas F\"osel, Petru Tighineanu, Talitha Weiss, and Florian Marquardt.
\newblock ``Reinforcement learning with neural networks for quantum feedback''.
\newblock \href{https://dx.doi.org/10.1103/PhysRevX.8.031084}{Phys. Rev. X {\bf
  8}, 031084}~(2018).
\newblock  \href{http://arxiv.org/abs/1802.05267}{arXiv:1802.05267}.

\bibitem{tishby2000information}
Naftali Tishby, Fernando~C. Pereira, and William Bialek.
\newblock ``The information bottleneck method''~(2000).
\newblock  \href{http://arxiv.org/abs/physics/0004057}{arXiv:physics/0004057}.

\bibitem{nguyen2017BMreview}
H.~Chau Nguyen, Riccardo Zecchina, and Johannes Berg.
\newblock ``Inverse statistical problems: from the inverse {Ising} problem to
  data science''.
\newblock \href{https://dx.doi.org/10.1080/00018732.2017.1341604}{Adv. Phys.
  {\bf 66}, 197--261}~(2017).
\newblock  \href{http://arxiv.org/abs/1702.01522}{arXiv:1702.01522}.

\bibitem{tramel2018tap}
Eric~W. Tramel, Marylou Gabri\'e, Andre Manoel, Francesco Caltagirone, and
  Florent Krzakala.
\newblock ``Deterministic and generalized framework for unsupervised learning
  with restricted {Boltzmann} machines''.
\newblock \href{https://dx.doi.org/10.1103/PhysRevX.8.041006}{Phys. Rev. X {\bf
  8}, 041006}~(2018).
\newblock  \href{http://arxiv.org/abs/1702.03260}{arXiv:1702.03260}.

\bibitem{Pozas2021RAPID}
Alejandro Pozas-Kerstjens, Gorka Mu{\~{n}}oz-Gil, Eloy Pi{\~{n}}ol,
  Miguel~{\'{A}}ngel Garc{\'{\i}}a-March, Antonio Ac{\'{\i}}n, Maciej
  Lewenstein, and Przemys{\l}aw~R Grzybowski.
\newblock ``Efficient training of energy-based models via spin-glass control''.
\newblock \href{https://dx.doi.org/10.1088/2632-2153/abe807}{Mach. Learn.: Sci.
  Technol. {\bf 2}, 025026}~(2021).
\newblock  \href{http://arxiv.org/abs/1910.01592}{arXiv:1910.01592}.

\bibitem{compApp}
Alejandro Pozas-Kerstjens, Senaida Hern\'andez-Santana, and David
  P\'erez-Garc\'{\i}a.
\newblock ``{Computational appendix of \textit{Physics solutions to machine
  learning privacy leaks}}''.
\newblock \href{https://dx.doi.org/10.5281/zenodo.6302728}{Zenodo {\bf
  6302728}, \phantom}~(2022).

\bibitem{dataset}
Global.health.
\newblock ``a data science initiative''.
\newblock \url{https://global.health}~(2021).
\newblock Accessed: 2021-03-22.

\bibitem{ateniese2015hacking}
Giuseppe Ateniese, Luigi~V. Mancini, Angelo Spognardi, Antonio Villani,
  Domenico Vitali, and Giovanni Felici.
\newblock ``Hacking smart machines with smarter ones: How to extract meaningful
  data from machine learning classifiers''.
\newblock \href{https://dx.doi.org/10.1504/IJSN.2015.071829}{Int. J. Secur.
  Netw. {\bf 10}, 137--150}~(2015).
\newblock  \href{http://arxiv.org/abs/1306.4447}{arXiv:1306.4447}.

\bibitem{shokri2017membership}
Reza Shokri, Marco Stronati, Congzheng Song, and Vitaly Shmatikov.
\newblock ``Membership inference attacks against machine learning models''.
\newblock In 2017 IEEE Symposium on Security and Privacy (SP).
\newblock \href{https://dx.doi.org/10.1109/SP.2017.41}{Pages 3--18}.
\newblock ~(2017).
\newblock  \href{http://arxiv.org/abs/1610.05820}{arXiv:1610.05820}.

\bibitem{davidMPS}
David P\'erez-Garc\'ia, Frank Verstraete, Michael~M. Wolf, and J.~Ignacio
  Cirac.
\newblock ``Matrix product state representations''.
\newblock \href{https://dx.doi.org/10.26421/QIC7.5-6-1}{Quantum Inf. Comput.
  {\bf 7}, 401--–430}~(2007).
\newblock
  \href{http://arxiv.org/abs/quant-ph/0608197}{arXiv:quant-ph/0608197}.

\bibitem{VidalMPS}
Guifr\'e Vidal.
\newblock ``Efficient classical simulation of slightly entangled quantum
  computations''.
\newblock \href{https://dx.doi.org/10.1103/PhysRevLett.91.147902}{Phys. Rev.
  Lett. {\bf 91}, 147902}~(2003).
\newblock
  \href{http://arxiv.org/abs/quant-ph/0301063}{arXiv:quant-ph/0301063}.

\bibitem{oseledets09}
Ivan~V. Oseledets.
\newblock ``A new tensor decomposition''.
\newblock \href{https://dx.doi.org/10.1134/S1064562409040115}{Dokl. Math. {\bf
  80}, 495--496}~(2009).

\bibitem{oseledets}
Ivan~V. Oseledets.
\newblock ``Tensor-train decomposition''.
\newblock \href{https://dx.doi.org/10.1137/090752286}{SIAM J. Sci. Comput. {\bf
  33}, 2295--2317}~(2011).

\bibitem{wahls2014}
Sander Wahls, Visa Koivunen, H.~Vincent Poor, and Michel Verhaegen.
\newblock ``Learning multidimensional {F}ourier series with tensor trains''.
\newblock In 2014 IEEE Global Conference on Signal and Information Processing
  (GlobalSIP).
\newblock \href{https://dx.doi.org/10.1109/GlobalSIP.2014.7032146}{Pages
  394--398}.
\newblock ~(2014).

\bibitem{chen2018}
Zhongming Chen, Kim Batselier, Johan A.~K. Suykens, and Ngai Wong.
\newblock ``Parallelized tensor train learning of polynomial classifiers''.
\newblock \href{https://dx.doi.org/10.1109/TNNLS.2017.2771264}{IEEE Trans.
  Neural Netw. Learn. Syst. {\bf 29}, 4621--4632}~(2018).
\newblock  \href{http://arxiv.org/abs/1612.06505}{arXiv:1612.06505}.

\bibitem{kargas2021}
Nikos Kargas and Nicholas~D. Sidiropoulos.
\newblock ``Supervised learning and canonical decomposition of multivariate
  functions''.
\newblock \href{https://dx.doi.org/10.1109/TSP.2021.3055000}{IEEE Trans. Signal
  Process. {\bf 69}, 1097--1107}~(2021).

\bibitem{wesel2021}
Frederiek Wesel and Kim Batselier.
\newblock ``Large-scale learning with fourier features and tensor
  decompositions''.
\newblock In
  \href{https://proceedings.neurips.cc/paper/2021/hash/92a08bf918f44ccd961477be30023da1-Abstract.html}{Advances
  in Neural Information Processing Systems}.
\newblock Volume~34, pages 17543--17554.
\newblock Curran Associates, Inc.~(2021).
\newblock  \href{http://arxiv.org/abs/2109.01545}{arXiv:2109.01545}.

\bibitem{gaugeBook}
Krishore~B. Marathe and Giovanni Martucci.
\newblock ``The mathematical foundations of gauge theories''.
\newblock North Holland Publishing Co. ~(1992).

\bibitem{oseledets2010decompo}
Ivan Oseledets and Eugene Tyrtyshnikov.
\newblock ``{TT}-cross approximation for multidimensional arrays''.
\newblock \href{https://dx.doi.org/10.1016/j.laa.2009.07.024}{Linear Algebra
  Appl. {\bf 432}, 70--88}~(2010).

\bibitem{tramer2016stealing}
Florian Tram{\`e}r, Fan Zhang, Ari Juels, Michael~K. Reiter, and Thomas
  Ristenpart.
\newblock ``Stealing machine learning models via prediction {APIs}''.
\newblock In
  \href{https://www.usenix.org/conference/\\usenixsecurity16/technical-sessions/presentation/tramer}{25th
  USENIX Security Symposium (USENIX Security 16)}.
\newblock Pages 601--618.
\newblock USENIX Association~(2016).
\newblock  \href{http://arxiv.org/abs/1609.02943}{arXiv:1609.02943}.

\bibitem{jagielski2020extraction}
Matthew Jagielski, Nicholas Carlini, David Berthelot, Alex Kurakin, and Nicolas
  Papernot.
\newblock ``High accuracy and high fidelity extraction of neural networks''.
\newblock In
  \href{https://www.usenix.org/conference/\\usenixsecurity20/presentation/jagielski}{29th
  USENIX Security Symposium (USENIX Security 20)}.
\newblock Pages 1345--1362.
\newblock USENIX Association~(2020).
\newblock  \href{http://arxiv.org/abs/1909.01838}{arXiv:1909.01838}.

\bibitem{molnarpeps}
Andr\'as Molnar, Jos\'e Garre-Rubio, David P\'erez-Garc\'ia, Norbert Schuch,
  and J.~Ignacio Cirac.
\newblock ``Normal projected entangled pair states generating the same state''.
\newblock \href{https://dx.doi.org/10.1088/1367-2630/aae9fa}{New J. Phys. {\bf
  20}, 113017}~(2018).
\newblock  \href{http://arxiv.org/abs/1804.04964}{arXiv:1804.04964}.

\bibitem{sidiropoulos2000}
Nicholas~D. Sidiropoulos and Rasmus Bro.
\newblock ``On the uniqueness of multilinear decomposition of {$N$}-way
  arrays''.
\newblock
  \href{https://dx.doi.org/10.1002/1099-128X(200005/06)14:3<229::AID-CEM587>3.0.CO;2-N}{J.
  Chemometrics {\bf 14}, 229--239}~(2000).

\bibitem{Liu_2019}
Ding Liu, Shi-Ju Ran, Peter Wittek, Cheng Peng, Ra\'ul
  Bl{\'{a}}zquez~Garc{\'{\i}}a, Gang Su, and Maciej Lewenstein.
\newblock ``Machine learning by unitary tensor network of hierarchical tree
  structure''.
\newblock \href{https://dx.doi.org/10.1088/1367-2630/ab31ef}{New J. Phys. {\bf
  21}, 073059}~(2019).
\newblock  \href{http://arxiv.org/abs/1710.04833}{arXiv:1710.04833}.

\bibitem{tensorLSTM}
Jiahao Su, Wonmin Byeon, Jean Kossaifi, Furong Huang, Jan Kautz, and Anima
  Anandkumar.
\newblock ``Convolutional tensor-train lstm for spatio-temporal learning''.
\newblock In
  \href{https://proceedings.neurips.cc/paper/2020/\\hash/9e1a36515d6704d7eb7a30d783400e5d-Abstract.html}{Advances
  in Neural Information Processing Systems}.
\newblock Volume~33, pages 13714--13726.
\newblock Curran Associates, Inc.~(2020).
\newblock  \href{http://arxiv.org/abs/2002.09131}{arXiv:2002.09131}.

\bibitem{tensortransformer}
Xindian Ma, Peng Zhang, Shuai Zhang, Nan Duan, Yuexian Hou, Dawei Song, and
  Ming Zhou.
\newblock ``A tensorized transformer for language modeling''.
\newblock In
  \href{https://proceedings.neurips.cc/paper/2019/hash/dc960c46c38bd16e953d97cdeefdbc68-Abstract.html}{Advances
  in Neural Information Processing Systems}.
\newblock Volume~32, pages 2232--2242.
\newblock Curran Associates Inc.~(2019).
\newblock  \href{http://arxiv.org/abs/1906.09777}{arXiv:1906.09777}.

\bibitem{Kuznetsov2019TRIP}
Maxim Kuznetsov, Daniil Polykovskiy, Dmitry~P Vetrov, and Alex Zhebrak.
\newblock ``A prior of a googol gaussians: a tensor ring induced prior for
  generative models''.
\newblock In
  \href{https://proceedings.neurips.cc/paper/2019/\\hash/4cb811134b9d39fc3104bd06ce75abad-Abstract.html}{Advances
  in Neural Information Processing Systems}.
\newblock Volume~32, pages 4102--4112.
\newblock Curran Associates, Inc.~(2019).
\newblock  \href{http://arxiv.org/abs/1910.13148}{arXiv:1910.13148}.

\bibitem{cheng2021PEPS}
Song Cheng, Lei Wang, and Pan Zhang.
\newblock ``Supervised learning with projected entangled pair states''.
\newblock \href{https://dx.doi.org/10.1103/PhysRevB.103.125117}{Phys. Rev. B
  {\bf 103}, 125117}~(2021).
\newblock  \href{http://arxiv.org/abs/2009.09932}{arXiv:2009.09932}.

\bibitem{barak2001obfuscation}
Boaz Barak, Oded Goldreich, Russell Impagliazzo, Steven Rudich, Amit Sahai,
  Salil Vadhan, and Ke~Yang.
\newblock ``On the (im)possibility of obfuscating programs''.
\newblock \href{https://dx.doi.org/10.1145/2160158.2160159}{J. ACM {\bf 59},
  1--48}~(2012).

\end{thebibliography}

\end{document}